\documentclass{article}

\usepackage[final]{neurips_2020}

\usepackage[utf8]{inputenc} 
\usepackage[T1]{fontenc}    
\usepackage{hyperref}       
\usepackage{url}            
\usepackage{booktabs}       
\usepackage{amsfonts}       
\usepackage{nicefrac}       
\usepackage{microtype}      

\title{Ideal theory in AI ethics}

\author{%
  Daniel Estrada\\
  Department of Humanities\\
  New Jersey Institute of Technology\\
  Newark, NJ 07102 \\
  \texttt{estrada@njit.edu} \\
  
}

\begin{document}

\maketitle

\begin{abstract}
This paper addresses the ways AI ethics research operates on an ideology of ideal theory, in the sense discussed by Mills (2005) and recently applied to AI ethics by Fazelpour \& Lipton (2020). I address the structural and methodological conditions that attract AI ethics researchers to ideal theorizing, and the consequences this approach has for the quality and future of our research community. Finally, I discuss the possibilities for a nonideal future in AI ethics.
\end{abstract}

\section{Resisting cynicism in AI ethics}

It’s easy to be cynical in AI ethics. My industry news streams are flooded with stories of appalling violations of security and privacy (Crawford et al. 2019), crisis opportunism in health care and education (Gerke et al. 2020, Teräs et al. 2020), monstrous applications of AI tech that trade in open racism, sexism, and violence (Birhane 2020, Ongweso Jr. 2020), applications that embed these technologies deep within the infrastructure of surveillance capitalism and the police state (Benjamin 2019b, Doctorow 2020), all wrapped in a form of deliberately wishful thinking that can only be built on an ignorance of history and the detachment of privilege (Moriah 2020). The few important victories achieved through activist organizing, such as bans on facial recognition in a few major cities (Nesterova 2020), have provided some needed relief and encouragement, but that relief is quickly overwhelmed by despair as the industry unfailingly sinks to new and devastating lows. 

And yet, the realities of tech activism on the front lines seem out of sync with AI ethics as a field of critical academic scholarship, which by all appearances is blossoming in these conditions. The major AI conferences have all created significant ethics programs, attracting a talented community of diverse voices and disciplinary backgrounds, with industry leaders offering considerable funding and support. Again, it’s easy to be cynical, but as a scholar it’s also hard to deny that this influence has been productive for the field. The quality and quantity of research associated with AI ethics (and related, more established fields like computer and robot ethics) has considerably matured and improved over the last five years. While the field is far from fully developed, we are slowly cultivating a canonical literature and focal case studies, and the roots of a scholarly discipline are taking shape. The field has largely overcome its early awkward obsession with superintelligence scenarios and Turing’s test, and has increasingly turned towards the impact that AI applications have on real people and communities. Importantly, researchers have started to take seriously the disproportionate impact these technologies have on different communities, and how these differences tend to reinforce systemic social biases in ways that put already marginalized persons at even greater risk. I do think AI ethics scholars deserve recognition for developing this research community in earnest from the earliest days of the recent boom in AI research. Other fields had to wait much longer to see similar developments.

At the same time, it is worth reflecting on the extent to which the successes of AI ethics scholars correlate with any positive impact on the industry we study. Have our policy recommendations, critical case studies, and theory-crafting sufficiently risen to meet the ethical challenges we face? For example, to what extent has the avalanche of fairness models in the literature actually shifted the ethical landscape towards justice? Are we here merely to produce an interesting and well-cited literature, or are we here to change industry and government policy for the better of us all? In some obvious ways these goals are in tension. Absent our dire ethical situation, the demand for these professional ethics communities (and the career opportunities they represent) are unlikely to be so generously supported. Of course this is not to accuse the community of a deliberate interest in perpetuating the current conditions of AI research, but simply to recognize the ways we all (myself included!) benefit from them, and moreover that these benefits are arranged by some of the very same industry players whose actions we are here to critique. Again, it’s easy to be cynical, but these realities are hard to ignore.

For the sake of a clean conscience, we might nevertheless attempt to resist cynicism and ask: how are we as AI ethics scholars to account for the simultaneous success and ineffectiveness of our field? The cynical, pessimistic response is to see these conferences as elaborate efforts in ethics washing, providing the appearance of ethical integrity and CSR through the cheap labor of young academics hungry for publishing credits, and with little power to achieve anything else. The cynical perspective has the considerable virtue of the evidence on its side. But in a collegial spirit of good faith, are there any non-cynical explanations available? The polar opposite of a cynical response would be an idealist or utopian response, one which argues (against all available evidence) that industry conditions are improving, that our scholarship is already having a clear and positive impact on professional standards and norms, that this is a political fight we are poised to win. Clearly, such a position is indefensible. But what else do we have? Should we just "give it time"?

This paper argues for a noncynical explanation: that AI ethics indulges in the ideology of ideal theory. By ideal theory, I do not mean the idealist or utopian perspective described above, which sees only light at the end of the tunnel. Although such utopian perspectives exist in the AI ethics literature, I don't think they are representative of the field or particularly responsible for the conditions we're in, so they aren't my concern here. I trust that my colleagues at this conference have at least some sense of our ethical predicament, and are not simply seeing the world through rose-colored glasses. Instead, I'm using ideal theory in the ideological sense developed by Charles Mills (2005), which describes a methodological tendency in ethical and political theory to rely on "idealization to the exclusion, or at least marginalization, of the actual" (p. 168). Unlike the utopian idealists, an ideal theorist may have a sincere commitment to ethics and justice, but this commitment finds expression through a preoccupation with abstractions and idealizations: with idealized models of society, of cognition, of governing institutions, and so on. While such idealizations can be useful in some contexts, an overemphasis on ideals can distract from more important or pressing concerns. As Khader (2018) says, “One defect of ideal theories… is their tendency to redirect our evaluative gazes to the wrong normative phenomena” (p. 36). According to Mills, ideal theory works to “abstract away from relations of structural domination, exploitation, coercion, and oppression” (p. 168), and so appeals to the "interests of the privileged who... have an experience that comes closest to that ideal, and so experience the least cognitive dissonance between it and reality" (p. 172).

In this sense and for these reasons, I argue that AI ethics researchers tend to engage the discourse in the mode of ideal theory, rather than addressing actual conditions of normative concern. This is not to accuse anyone in the community of a sinister motive, but simply to recognize how our collective interests align in such a way as to favor an idealized (and hence, politically and practically toothless) ethical discourse. Fazelpour \& Lipton (2020) laid the groundwork for an analysis of ideal theory in AI ethics at AAAI/AIES earlier this year with a fantastic case study of the fairness literature, a clear example of a reliance on ideals to the exclusion of the actual. The authors discuss the many ways the models of fairness explored in the literature systematically neglect injustices, distort harms, obscure an understanding of context, and provide little practical guidance in nonideal cases in the real world. However, the authors stop short of addressing the conditions that predispose AI ethics to the trappings of ideal theory in the first place, or what this disposition might say about the field.

Our goal for the rest of this paper is to understand why AI ethics researchers find the ideology of ideal theory so attractive. We will treat this task non-cynically in the sense that we will not be pessimistic or flippant about our capacity, as a scholarly community, to address and correct these weaknesses in our discourse. On the contrary, I think it is of the utmost importance that we take these challenges seriously and face these programmatic and methodological issues head-on. In the next section, I'll discuss some aspects of AI ethics scholarship that might help explain the appeal of ideal theory, and I'll point to some other cases worth considering for further analysis. I'll close with some words about a nonideal future for AI ethics. 

\section{Methodological challenges for AI ethics research}
In this section we'll discuss a few structural features of the AI ethics research community that incline our scholarly output towards the ideology of ideal theory. The list is by no means exhaustive, and is offered in no particular order except to get the discussion going. 

\subsection{Discursive instabilities} 
One way to understand the use of ideal theory in AI ethics is to notice how ideals can serve to abstract away from differences in methods, backgrounds, and training among AI ethics researchers. In other words, the appeal of ideal theory can partly be understood as a consequence of the interdisciplinarity of the field. Tech ethics has always been an interdisciplinary field \textit{par excellence}, an ethos that is undoubtedly a core strength of this field. One consequence of  interdisciplinarity amid a changing technical landscape is a persistent novicehood within the scholarly community. Even in expert settings like these professional conferences, nearly every participant (including very senior and prestigious people) will typically be utter novices on some important dimensions of any given discussion. It is quite common at these conferences to see scholars from all disciplinary backgrounds having only come recently to AI research, and who admit to having little technical insight into the algorithms motivating their work. In my experience it is even more common to see AI developers without any real background in ethical theory or technology studies expanding generously on the ethical and social dimensions of their work. In most cases these are positive and constructive contributions to the discourse! But such contributions also demand some humility and mindfulness of the constraints they place on the community. In a literature that is rapidly expanding and attracting interest from a broad background of scholars with no established standards, methods, or paths to expertise, the discourse can struggle to stabilize in productive or coherent ways. This lack of stability makes it difficult for the community to track the pace of news and technical innovation in AI, or to adequately prepare new generations of scholars for these challenges. 

To be clear, I do not mean to argue that AI ethics ought to place a greater emphasis on "expertise", or otherwise construct barriers to access and influence on the discourse. I think such gate keeping policies would be disastrous for the community. Instead, I want to point out how such discursive instabilities create conditions that favor the abstractions of ideal theory. Ideal theory as ideology offers ready ways to bridge epistemic and structural gaps between scholars in a highly interdisciplinary, highly technical, constantly changing, and well-funded field like AI ethics. In these conditions, the ideals function as conceptual anchors to a background ideology that artificially stabilizes the discourse and helps bridge its gaps. Put simply, it's relatively easy for two scholars from different backgrounds to agree that fairness is a virtue, and then cooperate on some implementation of the virtue. It's a lot harder to decide what fairness amounts to across a variety of real-world cases. By staying engaged with the abstractions, ideal theory enables scholarly cooperation at the level of ideology, and so for interdisciplinary scholarship to continue despite an ambiguous discourse. This point should not be confused with the suggestion that ideal theory enables more inclusive and interdisciplinary scholarship than its alternatives. On the contrary, the abstractions of ideal theory are able to bridge certain epistemic and structural gaps through a dynamic of ideological power and conformity that is often deliberately exclusive. Ideal theory tends to focus on ideals that reflect existing power structures and cultural assumptions. Thus, ideal theory not only operates as a means through which power articulates and reinforces its own interests; ideal theory can also be deployed by the powerless to demonstrate complicity with those ideals, either in an attempt to acquire power or to escape its wrath.

It would repeat the mistakes of ideal theory to respond to these discursive instabilities by demanding more rigorous or robust ideals, or by restricting epistemic access and participation. A nonideal alternative would turn our attention from the abstractions of theory to the actual conditions in which technologies are deployed and the people they impact. The few recent scholarly sources that have become foundational texts in AI ethics (Benjamin 2019a, Eubanks 2018, Noble 2018) are not aimed at an evaluation of ideals abstractly considered, but instead are focused squarely on the practical realities of our technological practices and the consequences these have for actual human lives. The lesson is that ideal theory is not the only way to stabilize a discourse. Centering the discourse on the human cost of AI is another resource for bridging disciplinary gaps, but one that does not turn on the comforts of abstraction and idealization.
    
\subsection{Systemic oppression} 
In addition to the scholarly constraints on AI ethics, the field also operates within and benefits from conditions of systemic oppression that disproportionately impact persons and communities already marginalized by racism and white supremacy, sexism and misogyny, homophobia, transphobia, ableism, xenophobic nationalism, and other forms of systemic violence, neglect, abuse, and exploitation (Birhane \& Guest, 2020). Mills describes how ideal theories "abstract away from relations of structural domination", and thereby serve the interests of those privileged few who benefit from the ideals and the structures of power they reinforce. While AI ethics draws from a diverse disciplinary background (and is improving, if slowly, on other diversity metrics), the resulting scholarship tends to operate within theoretical ideals that broadly serve the political and financial interests of the tech industry. AI ethics scholars tend to be tech enthusiasts to some degree, and rarely argue for tech prohibition outside of narrow contexts of use. Criticisms tend to stay focused on particular applications or algorithms, rather than the larger social context in which they operate. Engaging in ideal theory enables scholars to avoid the more uncomfortable incongruities of their work and the systemic injustices that make it possible. Why bite the hand that feeds us? 

As a result of this neglect, there is little consensus in the field on the tech policy crises of our time (eg, content moderation on social media, contact tracing apps, police and ICE contracts, etc.), and virtually no community procedures for developing such a consensus. Instead, our time is spent conjuring a thousand models of fairness and comparing their merits. Rather than flee to ideal theory and citation honeypots that marginalize the actual, a nonideal approach would encourage scholars to focus on the actual conditions of injustice faced by real people. Fazelpour and Lipton emphasize that "non-ideal theorizing about the demands of justice is a fact-sensitive exercise" (p. 10), and they advocate for empirically informed causal models for understanding the social dynamics of injustice. At the same time, they highlight the importance of "human thought, both to understand the social context and to make the relevant normative judgments" (p. 10). The upshot is that "fact-sensitivity" is not sufficient by itself, but requires a situated understanding informed by the communities directly involved in that context. We add to these suggestions the critical importance of recognizing the particular contexts of systemic oppression that both AI research and ethical oversight operate within. Insofar as ideal theory serves to deflect attention and criticism of these conditions, it also serves to perpetuate them.
    
\subsection{Capitalism and labor}
While the field of AI Ethics has made some progress in acknowledging conditions of systemic oppression in which tech is developed and used, the literature has struggled to face up to capitalism itself as the enabling condition of oppression, violence, and exploitation of marginalized peoples. Much has been made of the use of tech in police surveillance and the justice system, but much less attention has been paid to the labor conditions that enable the astronomical profits regularly reported in the industry. Much of the resistance to industry behavior has been coming from tech workers in the industry, who have been organizing and demonstrating effectively for years, despite repeated and high profile retaliation (Tarnoff 2020). Such issues are rarely taken up by AI ethics scholars, much less integrated into our policies and research goals. I argue that this neglect is enabled by an uncritical appeal to ideal theory. In the US especially there are deep layers of ideological and cultural resistance to a critique of capital as a corrupting force. This resistance is especially pernicious in AI ethics, where the most egregious ethical failures tend to have a financial motivation. Abdalla \& Abdalla (2020) have recently raised concerns about corporate influence over AI Ethics research and conferences, drawing a compelling analogy to Big Tobacco's manipulation of research on the medical consequences of smoking. Insofar as ideal theory serves to distract from the consequences of these conflicts of interest, it also serves to excuse and normalize them. A non-ideal alternative would put these structural realities at the center of our ethical discussions, as critical dimensions of the context in which AI ethics operates.

\section{A nonideal future for AI ethics}
The above influences are not comprehensive, but they conspire to set a low bar for achievement in AI ethics, and thus to ensure that scholars are satisfied with the limited analysis afforded by ideal theory. I submit that these influences point towards a non-cynical explanation for the simultaneous success and ineffectiveness of AI ethics scholarship. With Fazelpour \& Lipton, I likewise recommend a reorientation to nonideal theory that highlights the situated realities in which tech operates. Given the structural dispositions highlighted in this paper, we might apply a similar analysis of the use of ideal theory in AI ethics discussions around transparency, accountability, and explainability, all of which deploy ideals around justice and social ontology in analogous ways to the fairness literature. Such ideals are also commonly found in the literature on AI agency and autonomy, especially concerning autonomous vehicles and autonomous weapons systems. Finally, idealized social ontologies and hierarchies are clearly recognizable in debates around robot rights (Estrada 2020). In each of these cases, an abstract and idealized discourse can be challenged with nonideal alternatives that are sensitive to facts, situated contexts, and the lived experiences of actual people. Since this context sensitivity is likely to be central to any sophisticated ethical analysis, we should expect a non-ideal approach to play a central role if we are to make any progress towards justice. 

\section*{References}

AAAI (2019) Code of Ethics and Conduct. Retrieved from: https://aaai.org/Conferences/code-of-ethics-and-conduct.php

Abdalla, M., \& Abdalla, M. (2020) The Grey Hoodie Project: Big Tobacco, Big Tech, and the threat on academic integrity. \textit{Arxiv pre-print} Retrieved from: https://arxiv.org/abs/2009.13676 

Benjamin, R. (2019a). Race after technology: Abolitionist tools for the new Jim Code. Social Forces.

Benjamin, R. (Ed.). (2019b). Captivating technology: Race, carceral technoscience, and liberatory imagination in everyday life. Duke University Press.

Birhane, A., \& Guest, O. (2020). Towards decolonising computational sciences. \textit{ArXiv pre-print} Retrieved from: https://arxiv.org/abs/2009.14258

Birhane, A. [@Abebab] (2020, September 24) Every tech-evangelist: \#GPT3 provides deep nuanced viewpoint Me: GPT-3, generate a philosophical text about Ethiopia GPT-3 *spits out factually wrong and grossly racist text that portrays a tired and cliched Western perception of Ethiopia* 
(ht @vinayprabhu) [Tweet] Retrieved from: https://twitter.com/Abebab/status/1309137018404958215

Crawford, K., Dobbe, R., Dryer, T., Fried, G., Kaziunas, E., Kak, A., Mathur, V., McElroy, E., Sanchez, A., Raji, D., Rankin, J., Richardson, R., Schultz, J., West, S., \& Whittaker, M. (2019) AI Now Report 2019. AI Now Institute at New York University. Retrieved from: https://ainowinstitute.org/AI\_Now\_2019\_Report.html

Doctorow, C. (2020, August 26) How to destroy surveillance capitalism. OneZero. https://onezero.medium.com/how-to-destroy-surveillance-capitalism-8135e6744d59

Estrada, D. (2020). Human supremacy as posthuman risk. The Journal of Sociotechnical Critique, 1(1), 5.

Eubanks, V. (2018). Automating inequality: How high-tech tools profile, police, and punish the poor. St. Martin's Press.

Fazelpour, S., \& Lipton, Z. C. (2020). Algorithmic fairness from a non-ideal perspective. Proceedings of the AAAI/ACM conference on AI, ethics, and society (pp. 57–63).

Gerke, S., Shachar, C., Chai, P. R., \& Cohen, I. G. (2020). Regulatory, safety, and privacy concerns of home monitoring technologies during COVID-19. Nature medicine, 26(8), 1176-1182.

Khader, S. J. (2018). Decolonizing universalism: A transnational feminist ethic. Springer.

Mills, C. W. (2005). "Ideal theory" as ideology. Hypatia, 20(3), 165–184.

Moriah, C. [@caitlinmoriah] (2020, September 27) I just watched the Social Dilemma.... oof, okay. All other issues aside, I cannot stop thinking about this throwaway comment about bicycles by Tristan Harris: [Tweet] Retrieved from: https://twitter.com/caitlinmoriah/status/1310312499355422720

National Society of Professional Engineers. (2019). NSPE Code of ethics for engineers. Retrieved from: https://www.nspe.org/resources/ethics/code-ethics

Nesterova, I. (2020). Mass data gathering and surveillance: the fight against facial recognition technology in the globalized world. In SHS Web of Conferences (Vol. 74, p. 03006). EDP Sciences.

Noble, S. U. (2018). Algorithms of Oppression: How Search Engines Reinforce Racism. NYU Press.

Ongweso Jr, E. (2020, September 28). An AI Paper Published in a Major Journal Dabbles in Phrenology. Vice Motherboard. https://www.vice.com/en/article/g5pawq/an-ai-paper-published-in-a-major-journal-dabbles-in-phrenology

Tarnoff, B. (2020, May 4) The Making of the Tech Worker Movement. Logic Magazine. Retrieved from:  https://logicmag.io/the-making-of-the-tech-worker-movement/full-text/

Teräs, M., Suoranta, J., Teräs, H., \& Curcher, M. (2020). Post-Covid-19 education and education technology ‘solutionism’: A seller’s market. Postdigital Science and Education, 1-16.

\end{document}